\documentclass[10pt]{iopart}
\usepackage{bm}
\usepackage{subfigure}
\usepackage{enumerate, pdflscape, graphicx}
\usepackage{pgfplots,pgf,tikz} \usetikzlibrary{arrows}
\usepackage{booktabs, hyperref}

\begin{document}
\title[Serious Game]{Coordinating vector field equations and diagrams with a serious game in introductory physics}
\author{P Klein$^1$, N Burkard$^2$,  L Hahn$^1$, M N Dahlkemper$^1$, K Eberle$^2$, T Jaeger$^2$, J Kuhn$^3$ and M Herrlich$^2$}
\address{$^1$Department of Physics, Physics Education Research, Georg-August-University G\"{o}ttingen, Friedrich-Hund-Platz 1, 37077 G\"{o}ttingen, Germany}
\address{$^2$Department of Electrical and Computer Engineering, Serious Games Engineering, Technische Universit\"{a}t Kaiserslautern, Paul-Ehrlich-Straße 11, 67663 Kaiserslautern, Germany}
\address{$^3$Department of Physics, Physics Education Research, Technische Universit\"{a}t Kaiserslautern, Erwin-Schr\"odinger-Str. 46, 67663 Kaiserslautern, Germany}

\ead{pascal.klein@uni-goettingen.de}
\bibliographystyle{unsrt}
\begin{abstract}
Mathematical reasoning with algebraic and graphical representations is essential for success in physics courses. Many problems require students to fluently move between algebraic and graphical representations. We developed a freely available serious game to challenge the representational fluency of introductory students regarding vector fields. Within the game, interactive puzzles are solved using different types of vector fields that must be configured with the correct mathematical parameters. A reward system implemented in the game prevents from using trial-and-error approaches and instead encourages the player to establish a mental connection between the graphical representation of the vector field and the (algebraic) equation before taking any action. For correct solutions, the player receives points and can unlock further levels. We report about the aim of the game from an educational perspective, describe potential learning scenarios and reflect about a first attempt to use the game in the classroom.
\end{abstract}


\submitto{\EJP}
%
\maketitle
%

\section{Introduction}
\begin{figure}[b!]
\centering
\includegraphics[width=0.45\linewidth]{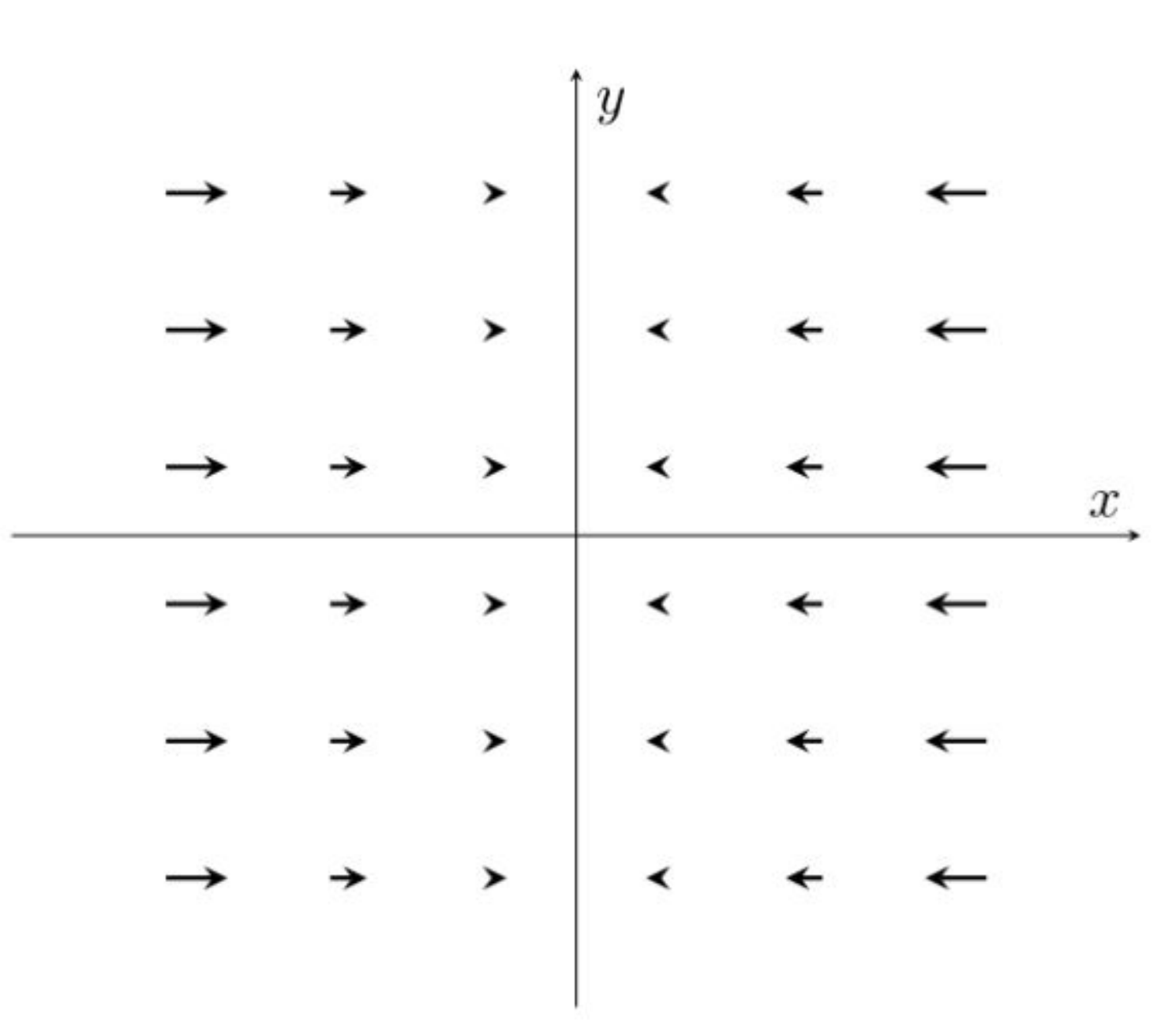} ~~\includegraphics[width=0.45\linewidth]{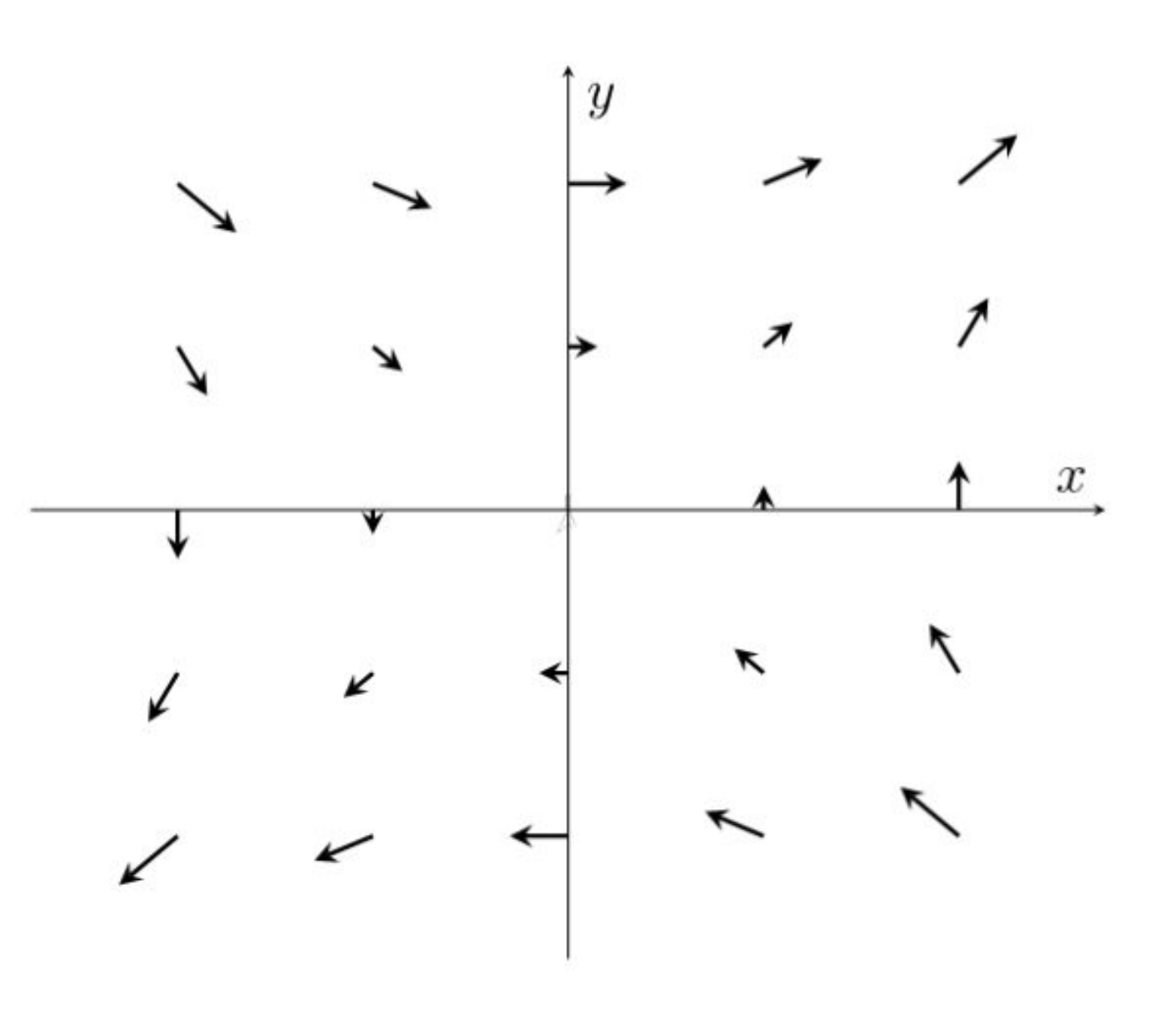}
\caption{Two examples of 2D vector field diagrams. The left vector field is described by the equation $\boldsymbol{F}(x,y)=-x ~\boldsymbol{\hat{x}}$, and the right one by $\boldsymbol{F}(x,y)=y ~\boldsymbol{\hat{x}} + x ~\boldsymbol{\hat{y}}$}
\label{fig:mc}
\end{figure}

Vector fields are mathematical objects that assign a vector to every point in the space or in a subset of space (e.g., the two-dimensional plane). Vector fields are important in many physics subjects, some of which are part of the introductory physics curriculum. Examples include Newton's gravitational field, velocity fields of fluids, and electromagnetic fields. Two illustrative examples of vector fields are given in Fig.~\ref{fig:mc}.

There are decades of research about students' difficulties associated with vectors \cite{Knight1995, Nguyen2003, Flores2004, VanDeventer2007, Barniol2014, Susac2018, Carli2020}. Numerous studies have shown that first-year university students have substantial difficulties regarding basic vector concepts, e.g., interpreting graphical properties such as direction, length and component decomposition, vector addition and subtraction, or scalar and cross products \cite{Knight1995, Nguyen2003, Flores2004}. Based on the broad body of research, a reliable instrument to assess students' understanding of vectors was developed, validated, and used by independent research groups \cite{Barniol2014, Susac2018, Liu2019}. Students' difficulties occur both with and without physics contexts \cite{VanDeventer2007, Carli2020}, i.e., when vector concepts are treated purely mathematically. A comprehensive knowledge of these concepts is a prerequisite for understanding vector fields, thus, students' difficulties with single vectors have a direct impact on their understanding of vector fields. In addition, students might understand basic concepts for single vectors but show difficulties applying the concepts to vector fields \cite{Bollen17}. For instance, students might succeed to superimpose two force vectors but they might fail to superimpose two force vector fields. 

\begin{figure}[b!]
    \centering
 \includegraphics[width=0.85\linewidth]{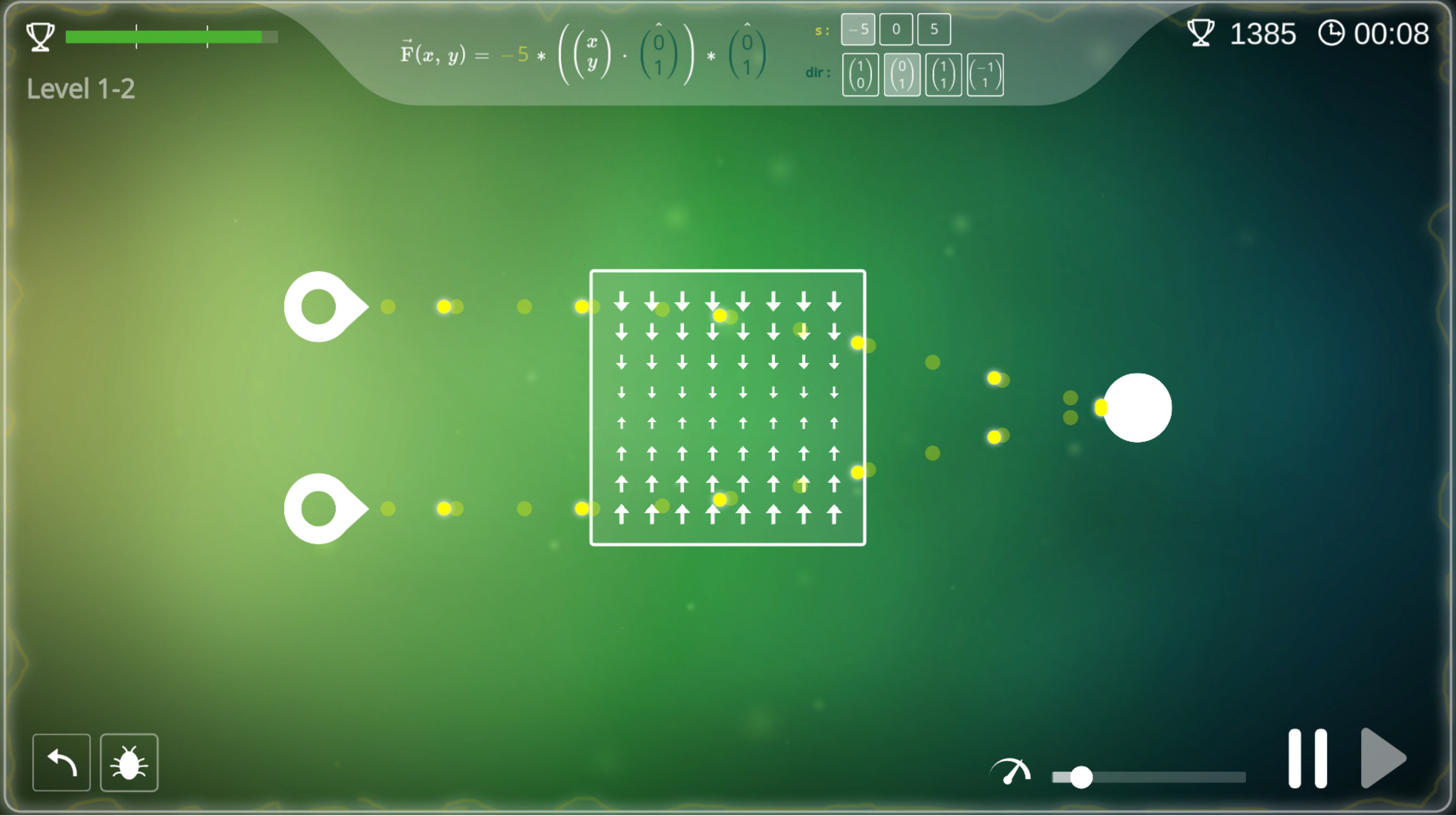}
    \caption{The effect of an appropriate vector field on the particles: The particles (yellow dots) move from the sources (left) to the target (right). The vector field equation and the parameters are present on the upper part of the screen.}
    \label{fig:principle}
\end{figure}

Physics concepts, such as conservation laws or cause-effect relationships, are usually expressed using (one or more) external representations \cite{DeCock2012}. Common representations of vector fields are both graphical representations, e.g. vector field diagrams (see Fig.~\ref{fig:mc}), and algebraic expressions using unit vectors in a previously defined coordinate system (see caption of Fig. \ref{fig:mc}) \cite{Bollen17}. The coordination of both forms of representations (diagram and algebraic expression) represents a further challenge for students in general \cite{Nieminen2012} and, particularly, for vector fields \cite{Bollen17}. Among others, the study of Bollen \textit{et al.} suggest that a confident and flexible handling of multiple representations can have a positive impact on learning and problem solving and to develop domain-specific expertise (see Sect. \ref{sec:educ} for more details). Indeed, several of the aforementioned learning difficulties regarding vectors were found to originate from a lack of understanding of the connections between the algebraic and geometric aspects of vectors \cite{Liu2019}. The serious game presented here is intended to foster the skills and routines for changing between representations in the context of vector fields that could complement traditional instructions. The aim of the Vector Field Game consists of guiding moving particles from a source to a target by choosing appropriate vector fields. The player can manipulate the algebraic expression and directly observe the impact on the field representation and on the movement of the particles, see. Fig. \ref{fig:principle}

\section{Serious games in physics education}

Games and playing in general as forms of learning pose an important factor in human development, especially, in early years as a child.  Play and games follow rules, at least given implicit but often fully formalized. There have been early (non-digital) examples of the latter for educational purposes especially in the military domain, e.g., GO and Chess, to train tactical and strategic thinking and decision making.

The term \textit{serious games} was first described by Abt in 1970 as ``games [that] have an explicit and carefully thought-out educational purpose and are not intended to be played primarily for amusement''\cite{Abt1987} originally not referring to digital games. The term itself is a bit contradictory, since it might imply that other games might not be serious or that serious games should not be entertaining. There are many other terms used for specific domains (games for learning, edutainment, educational games, etc.). However, the term serious game is often used as the best available summarizing term that encompassed many different approaches and domains~\cite{Wilk2016}.

With the ongoing digitalization and transformation of society, digital games have entered the scene and with the cultural and economic success of (entertainment) games~\cite{Murr2017}, a whole generation has been socialized by gaming which in turn may contribute to the success of serious games~\cite{Wilk2016}.

The research on serious games in physics education dates back to the work of White in 1986 who designed computer games to teach Newton’s laws of motion \cite{White}. In a controlled study she found a significant effect of playing the game on the students’ conceptional understanding of Newtonian dynamics, especially about circular motion. The positive effect was explained by the fact that the students make connections between their intuitive beliefs about physics and their knowledge about formal physics, encompassing, e.g., vector addition. Students who lack an understanding of Newton’s laws can rely on general problem-solving heuristics combined with feedback to reach the goal. White argues that the serious game should not require an a priori knowledge of physics, since “students might find the game dull if they already knew the physics and impossible if they did not” \cite[p. 73]{White}. The research on the learning outcome of serious games in general is summarised by meta-analyses such as the one by Clark et al. \cite{Clark} which found positive effects of game-based learning and was in accordance with several other meta-analyses carried out before \cite{Vogel, Sitzmann, Wouters}. In the meta-analysis by Li et al. three studies on game-based learning in the field of physics with undergraduate students were documented \cite{Anderson, Johnson, Mayer}. In these studies, the games "Supercharged!" and "Circuit game" are examined. The first is used to teach electrostatics while the second is a puzzle game to learn how electrical circuits work. Furthermore, Eccheverría et al. proposed a structured methodology for game design for the conceptual understanding of physics \cite{Eccheverria}. The three principles include that the independent variables of the simulation are integrated as low-level game atoms, the dependent variable of the simulation is integrated as a scaffolded game atom, and this connects to the goal through an additional game atom that provides an interesting challenge. Our approach in the Vector Field Game followed basically the work of White and considers also the structure of Eccheverría et al., see. Sect. \ref{sec:design}

Most of the studies which have been carried out on serious games in physics education where on primary and secondary level, while on the higher education level research focussed on simulations and virtual lab environments (e.g., \cite{Darrah}). While most of the games have been designed for the computer, advances in technological development implicate an extension to gaming consoles and especially handheld devices. In this context, a commercial video game for the PlayStation 3 console was used to teach kinematics to undergraduate students \cite{Mohanti}. The game presented here is available for PC, smartphones, and tablet computers.

\section{Educational and theoretical background: Handling multiple representations of vector fields} \label{sec:educ}

Learning processes with multiple representations are a central subject of physics education research. Ainsworth formulated three central functions of multiple representations (MR) that can facilitate learning \cite{Ainsworth1999}; (1) MR contain complementary information, i.e.\ the learner benefits from the advantages from both representations \cite{Seufert2003}; (2) MR can help the learner to develop a better understanding of the subject by using one representation to limit possible misinterpretations by another; and (3) MR can help the learners to develop a deeper understanding of a concept. Based on these functions, many researchers report a positive effect of using MR on knowledge acquisition and problem-solving skills \cite{Dufresne1997, Even1998, Nieminen2012,Rau2009, Rosengrant2007, VanHeuvelen1991, VanHeuvelen2001, Klein2018, Becker2020}. For example, Nieminen, Savinainen and Viiri (2012) found a strong correlation between learners' ability to interpret multiple representations consistently (representational consistency) and their learning gain in a study on forces \cite{Nieminen2012}. Their result confirms that judicious use of multiple representations can contribute to a better understanding of physical concepts. In the context of visually interpreting vector field concepts (e.g., the flux concept), Klein et al. found that students’ performance improved when two different strategies were introduced instead of only one strategy, and students performed best when they were free to choose between the two strategies \cite{Klein2018}. Their finding supports the idea of introducing multiple representations to foster student understanding. 
In general, the understanding that learners gain by using multiple representations is broader, deeper\cite{Even1998} more flexible and robust \cite{Ainsworth2002} than that which they generate using single representations. 

To sum it up, the use of multiple representations allows problems and concepts to be considered from different perspectives. The ability to use different strategies can have synergistic effects on the construction of coherent knowledge structures \cite{Seufert2003} and might help to evoke varied mental pictures assisting to make new discoveries. Feynman who became famous for inventing new representations to describe complex processes with greater simplicity put the aforementioned arguments in a nutshell  \cite{Feynman}:
\begin{quote}
\textit{``Psychologically they [representations of physical concepts] are different because they are completely inequivalent when you are trying to guess new laws".}
\end{quote} 

Despite the potential benefits described above, there are numerous studies showing that the use of multiple representations does not per se lead to higher learning outcomes \cite{Rau2009, Ainsworth2008, Yerushalmy1991}. In order to benefit from the advantages of multiple representations in learning and problem solving, a deeper understanding of the representations is necessary, which is described with the help of two competences, following De Cook (2012) and Nistal et al (2009, 2012) \cite{DeCock2012, Nistal2009, Nistal2012}:
\begin{enumerate}
\item representational fluency: This competence enables the interpretation and construction of representations and enables the correct and quick switching and translation between different forms of representation \cite{Bieda2009}.
\item representational flexibility: this competence enables the choice of an appropriate form of representation in a given problem or learning situation and involves the ability to take into account characteristics of the subjects interacting with the representation and the context of the interaction \cite{Nistal2009, Nistal2012}.
\end{enumerate}
A central aspect of representational fluency is called representational competence and includes the knowledge of how representations are to be interpreted and how they represent information about the learning content \cite{DeCock2012, Rau2017}. If learners do not know how the visual representation encodes information (also referred to as visual understanding~\cite{Rau2017}), multiple representations can have negative effects on problem-solving ability and learning success \cite{DeCock2012,Yerushalmy1991}.
 
 Bollen et al (2017) investigated students' errors when switching representations between vector field diagrams, field line diagrams and algebraic expression \cite{Bollen17}. 
When constructing a vector field diagram using an algebraic formula expression, errors were found that were mainly due to problems with vector addition, representing the change in length and direction of the vectors with increasing distance from the origin, and, for the reverse direction (i.e., changing from graphical to algebraic representations), the choice of a suitable coordinate system and the use of unit vectors posed a problem. Gire and Price (2012) who dealt with representation changes in the context of electric and magnetic fields also found that students often had problems differentiating between coordinates and components \cite{Gire2012}. This was especially the case with ``mixed'' dependencies, e.g. when the $x$-component was a function of the $y$-coordinate.  

Against the background of these particular findings, the authors call for specific instructions to train representational fluency in dealing with vector fields \cite{Bollen17,Gire2012}, to which the Vector Field Game can contribute by playfully practicing the transition between vector field diagram and formula. 

\section{The Vector field game: game concept and design}

\begin{figure}
    \centering
    \includegraphics[width=\linewidth]{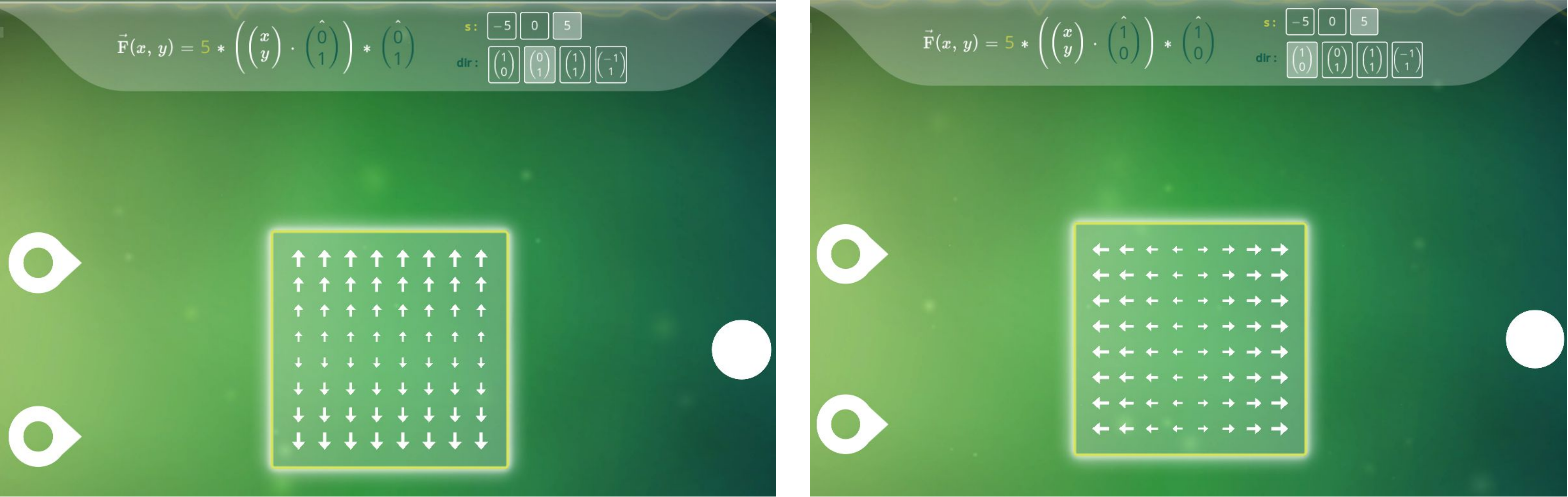} 
    \caption{In-game coordination between equation and vector field. The diagram changes based on the parameter of the field}
    \label{fig:fluency}
\end{figure}

The Vector Field Game was designed to complement teaching of vector fields and to teach intuition about the abstract concept in an interactive fashion. Therefore, we chose a casual style game to improve accessibility and to visualize the direct effect of different vector fields on particles. 
The game was developed in an iterative design process as proposed by Fullerton~\cite{Full2014}. This enabled us to focus on player experience (``playcentric design process'') in every development phase.

The game itself consists of recurring elements (see Fig.~\ref{fig:vgv}):
In every level, there is at least one source of particles (drop shaped), one goal (circle), and one but usually multiple boxes representing vector fields. The player needs to select the appropriate vector fields in order to guide the particles to the goal and solve the level. In the design framework of Eccheverría et al. \cite{Eccheverria}, the vector field and the equation correspond to the low-level game atoms (independent variable that can be controlled by the user), and the track of the particle is the dependent variable. The challenge is to establish a connection between the source of the particle and the target which reflects an additional game atom.

The scaffolds that Eccheverría et al. describes are also present in the game  as two types of boxes. Boxes with light background indicate a fixed vector field type whose parameters must be chosen by clicking on the box and subsequently selecting them on the top bar (see Fig.~\ref{fig:vgv}, right). Boxes with grey background indicate not a specific field type, so the player has to choose the correct type by dragging and dropping the respective icon from the button bar in the desktop version. The mobile version opens an extra menu in which the type can be selected by tapping.
The top bar is only present when a box is selected (see Fig.~\ref{fig:principle}) and always displays the vector field formula (as link to the mathematical representation) and possible parameters for it. The game includes four kinds of vector fields: constant vector fields, hook fields, radial fields and rotational fields. When changing the parameters of the vector field, the arrow visualization in the vector field box is instantly updated so that the player can visually explore the link between the vector field formula and the arrow representation and thus encouraging representational fluency.

By selecting the start button on the lower right, the game starts the particle simulation (see Fig.~\ref{fig:fluency}). The particles leave ghost spots in order to trace their paths which is very useful to identify mistakes in the vector field selection. The simulation speed can as well be chosen with the slider left of it. If all particles reach their destination, the level is finished and the player is rewarded with a summarizing screen providing an additional ``star rating'' based on score. From here, the player can continue or return to the main screen.
To raise the difficulty and make it more interesting, levels may also require the combination of more than one vector field or the levels include obstacles. To increase the difficulty further,  sources may spawn ``charged''
particles, denoted by a +/- sign. "Positively charged" particles are yellow color coded, "negatively charged" ones red. Red particles interact with the field in opposite direction. Please note, that the vector fields have no physical meaning, so no physical meaning in charge is intended here.

On the top right of the screen, a timer counting upwards and the current score are displayed, on the left the numbering of the current level and display representing the current rating where each part represents a star.

\subsection{Design principles}\label{sec:design}

In order to obtain acceptable usability, which we identified as a key quality requirement, we included a tutorial and focused on the development of an intuitive interface.
Our goal is to facilitate the acquisition of intuition complementing the theoretical knowledge on vector fields. Therefore, we refrained from using unconstrained mathematics as it would offer too many degrees of freedom without an additional benefit in terms of intuition as minimal changes of parameters would almost be indiscernible but might lead to unintended results in some cases. However, we also wanted to avoid pure trial and error type play, which was a difficult balance to achieve, and partly, this is still open for evaluation.

\begin{figure}[t]
\centering
\includegraphics[width=0.45\linewidth]{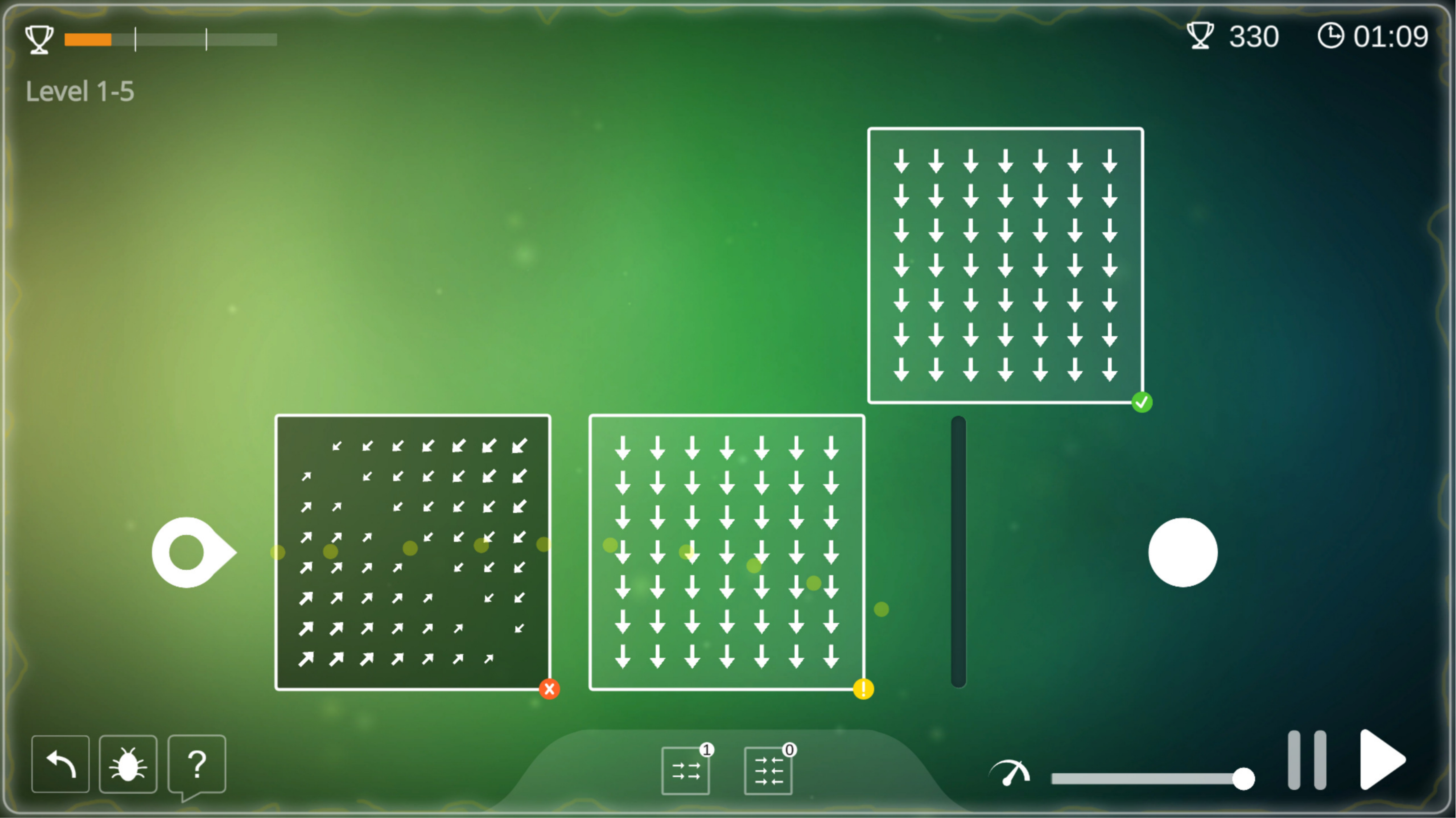}~~\includegraphics[width=0.45\linewidth]{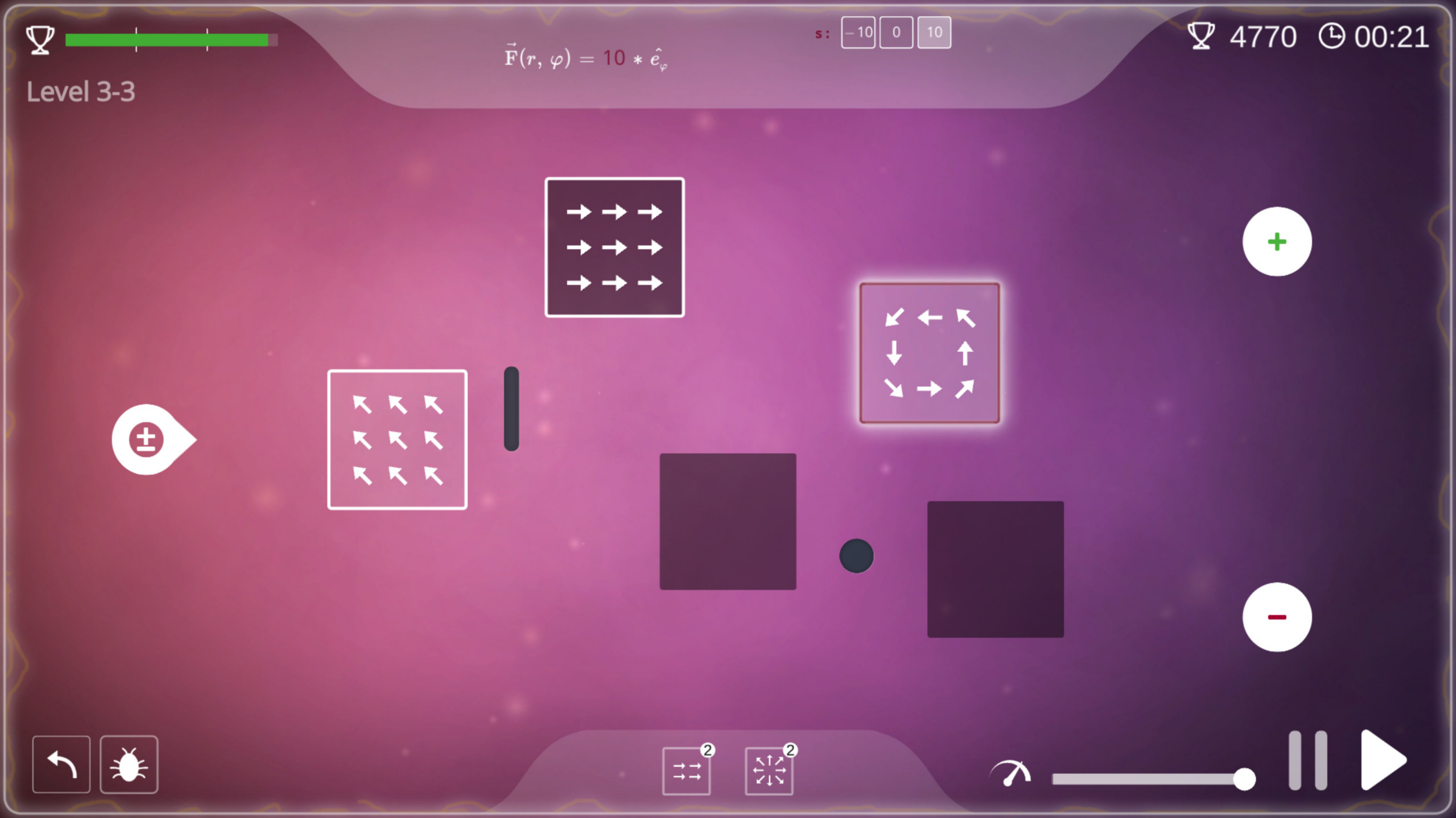}
\caption{In game view of two different levels of the Vector Field Game. Left: The simulation started but the particles did not enter the goal. The colored icons on the bottom right of each vector field box denote whether the correct type was used. Right: A level with ``charged'' particles. The positive and negative particles have each have a respective goal.}
\label{fig:vgv}
\end{figure}

We followed an iterative design and development process to find a good balance between offering enough degrees of freedom in the formulas and different types of vector fields -- to facilitate actually making errors to learn from -- but also provide guidance and scaffolding in the form of different design elements. First of all, the level design itself and the progression of levels is set up to start simple and raise complexity slowly: the game introduces simpler vector field types first, with fewer parameters to choose from, and the early levels include overall fewer vector fields.
In addition, we designed a graphical indicator as a direct reaction to player testing providing feedback in different tiers: red for wrong field type, amber for correct field type but wrong parameters, and green for correct field type with correct parameters (see Fig.~\ref{fig:vgv}).
As countermeasure for blindly trying to solve the level by random choices, the number of parameter selections and the overall level time influence the resulting score. We decided to put a mild time pressure on players to provide a motivating challenge and to avoid an overly slow and boring pace of the game. However, every level is still solvable even if the time runs out to reduce player frustration in challenging levels.

In general, the reward design to motivate the players is twofold: on one hand, intrinsic motivation is boosted by good usability and stream-lined, non-distracting interaction as well as graphical aesthetic design. On the other hand, extrinsic motivation is given in the form of points and star ratings for each level. The later also facilitates replayability because the star ratings are also shown in an overview screen of all levels unlocked so far to motivate players to beat their own high-scores.

\subsection{Resources}

The game is free and available on the desktop\footnote[1]{\url{https://tuk-software.procampus.de/de/vektorfeldspiel/}} for Windows and macOS and on mobile platforms for Android\footnote[2]{\url{https://play.google.com/store/apps/details?id=de.unikl.eit.sge.vectorfieldgame.android}} and iOS\footnote[3]{\url{https://apps.apple.com/de/app/vektorfeldspiel/id1517002472}}.

\subsection{Vector fields used in the game}

Note that the vector fields that were constructed for the serious game and for the educational studies do not necessarily have a physical meaning. For our aims we wanted to avoid that lacking physical knowledge hinders the acquisition of representational fluency. To become knowledgeable physicists, however, students definitely must supplement their mathematical skills with conceptual knowledge of the physical world. For instance, the left vector field in Fig. \ref{fig:mc} could be interpreted as the force field of an one-dimensional spring, ${F}(x)=-k~x$, that was extended to the two-dimensional plane, $\boldsymbol{F}(x,y)=-k~x~\boldsymbol{\hat{x}}$. Even though conceptual knowledge in physics of this kind might be helpful for solving the puzzles, they are no prerequisite.

\section{First distribution among students and research potential}
\begin{figure}
    \centering
    \includegraphics[width=0.9\linewidth]{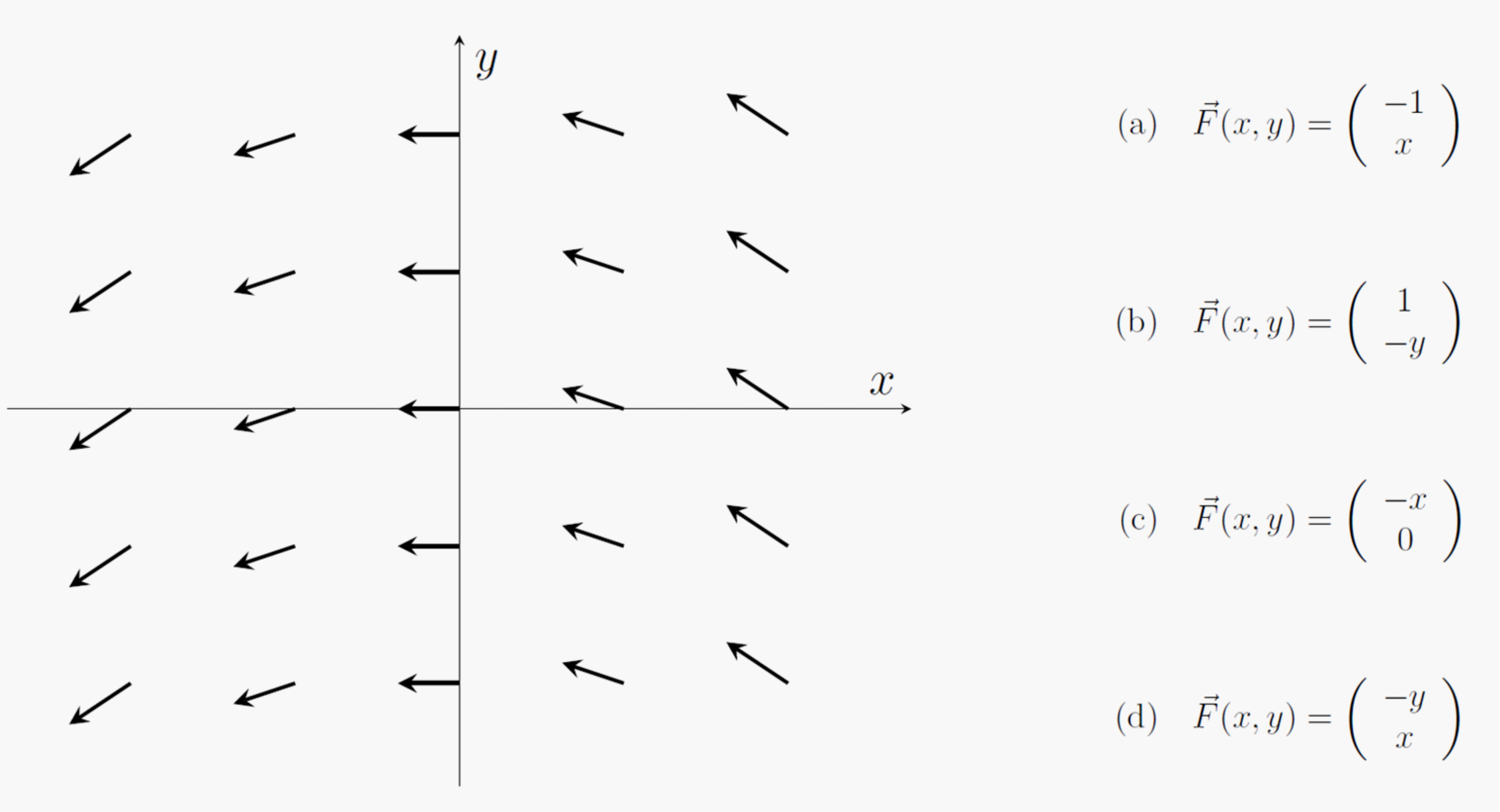}
    \caption{One out of eight multiple choice items to assess students' representational fluency}
    \label{fig:testitem}
\end{figure}

In January 2020, the game was distributed to first-term physics students and an investigation of the presumed learning potential of the game was prepared. For this purpose, a diagnostic test was  designed that addresses the connection between the mathematical representation form (formula and equation) and the corresponding graphical representation of vector fields - a competence that should be promoted by the game from an educational point of view, see Sect. \ref{sec:educ}. The instrument includes 8 vector field diagrams in total, and students were asked to select the corresponding vector field equation out of four alternatives (see Fig. \ref{fig:testitem} for an example). The average ability of the students ($N=68$) to establish coherence between formula and diagram determined by the test instrument was 57.3\% after correction for guessing, measured as pre-test, i.e. without playing the game. 

 Initially it turned out that the informal distribution of the game without instructions did not produce the desired response. Of the 68 students to whom the download link for the game was issued with the request to play the game during the upcoming four weeks, only 26 actually downloaded the game. Of those who downloaded the game, 13 students were engaged with the game for less than 10 minutes, 9 between 10 minutes and 1 hour, and 6 students said they had played the game to the end. 62\% of the students thought that the game was fun and the same number said that they learned something from it.
 Due to the low student participation rate, a post-test was not carried out to reassess the competence, so no conclusions can be drawn about the change in this competence that the game might cause. In follow-up examinations it is advisable to anchor the game formally and instructionally. In order to lower the usage threshold for students in the future and to enable testing on a larger scale, the porting and publication of the game to mobile platforms (Android and iOS) was carried out. 

\section{Concluding remarks and outlook}
In times of increasing importance of online learning due to the COVID-19 pandemic, educational technology and digital teaching materials gain in value.  We reported about  a freely available serious game that offers a contribution for physics students and lectures. The Vector field Game was developed to challenge the representational fluency of introductory students regarding vector fields. It was discussed that (1) being fluent to make connections between equations and diagrams is an important skill in mathematics and physics education, particularly in the context of vector fields, and (2) that serious games can have the potential to substantially foster learning by a mixture of using intuitions, problem-solving heuristics, and feedback from the game. Combining both lines of research, i.e., learning with multiple representations and serious games engineering, we suppose that the Vector Field Game is beneficial for introductory physics students. The actual impact on student learning has to be tested.

\section*{Acknowledgements}
The project was partly funded by the "LehrePlus"-grant of the TU Kaiserslautern, aiming at sustaining and improving the quality of teaching and learning in university education.

\end{document}